 \documentclass[final,5p,times,twocolumn]{elsarticle}
 \biboptions{comma,sort&compress}
\usepackage{here}
 \usepackage{graphicx}
 \usepackage{epsfig,slashed}

\usepackage{amssymb}
 \usepackage{amsthm}

\def\beq{\begin{equation}}
\def\eeq{\end{equation}}
\def\bea{\begin{eqnarray}}
\def\eea{\end{eqnarray}}
\def\bq{\begin{quote}}
\def\eq{\end{quote}}

\def\lb{\label}
\def\nnb{\nonumber}
\def\ga{\left(}
\def\dr{\right)}

\def\rar{\rightarrow}

\def\nnb{\nonumber}
\def\la{\langle}
\def\ra{\rangle}
\def\nin{\noindent}
\def\ba{\vspace*{-0.2cm}\begin{array}}
\def\ea{\end{array}\vspace*{-0.2cm}}

\def\b{$\bullet~$}

\def\als{\alpha_s}

\def\gg2{ \la\alpha_s G^2 \ra}
\def\gg3{g^3f_{abc}\la G^aG^bG^c \ra}
\def\ggg4{\la\als^2G^4\ra}

\def\ss{\la\bar{s}s\ra}

\journal{Physics Letters B}

\begin{document}

\begin{frontmatter}



\title{Mass-splittings of doubly heavy baryons in QCD}


 \author[label1,label2]{R.M. Albuquerque\fnref{label3}}
  \address[label1]{
Instituto de F\'{\i}sica, Universidade de S\~{a}o Paulo, 
C.P. 66318, 05389-970 S\~{a}o Paulo, SP, Brazil.}
\fntext[label3]{FAPESP thesis fellow within the France-Brazil bilateral exchange program.}
\ead{rma@if.usp.br}

 \author[label2]{S. Narison\corref{cor1}}
  \address[label2]{Laboratoire
de Physique Th\'eorique et Astroparticules, CNRS-IN2P3 \& Universit\'e
de Montpellier II, 
\\
Case 070, Place Eug\`ene
Bataillon, 34095 - Montpellier Cedex 05, France.}
\cortext[cor1]{Corresponding author}
\ead{snarison@yahoo.fr}



\begin{abstract}
\noindent
We consider (for the first time) the ratios of  doubly heavy baryon masses (spin 3/2 over spin 1/2 and SU(3) mass-splittings) using double ratios of sum rules (DRSR), which are more accurate than the usual simple ratios often used in the literature for getting the hadron masses.
In general, our results agree and compete in precision with potential model predictions. In our approach, the $\alpha_s$ corrections induced by the anomalous dimensions of the correlators are the main sources of the $\Xi^*_{QQ}- \Xi_{QQ}$ mass-splittings, which seem to indicate a $1/M_Q$ behaviour and
can only allow the electromagnetic decay $\Xi^*_{QQ}\to\Xi_{QQ}+\gamma$ but not to ${\Xi_{QQ }}+ \pi$. 
Our results also show that the SU(3) mass-splittings are (almost) independent of the spin of the baryons and behave approximately like $1/M_Q$, which could be understood from the QCD expressions of the corresponding two-point correlator. Our results  can improved by including radiative corrections to the SU(3) breaking terms and can be tested, in the near future, at Tevatron and LHCb.
\end{abstract}


\end{frontmatter}


\section{Introduction}
\nin
In a previous paper \cite{HBARYON}, we have considered, using double ratios\,\cite{SNGh,SNhl,SNFBS,SNFORM,SNme+e-} of QCD spectral sum rules (QSSR) \cite{SVZ,SNB} (DRSR), the splittings due to SU(3) breakings of the baryons made with one heavy quark. In this paper, we pursue this project  in the case of doubly heavy baryons. 
The absolute values of the doubly heavy baryon masses of spin 1/2 $(\Xi_{QQ}\equiv QQu)$ and spin 3/2 $(\Xi^*_{QQ}\equiv QQu)$ have been obtained
using QCD spectral sum rules (QSSR) (for the first time) in \cite{BAGAN} with the results in GeV:
\bea
M_{\Xi^*_{cc}}(3/2)&=& 3.58(5)~,~~~~M_{\Xi^*_{bb}}(3/2)=10.33(1.09)~,\nnb\\
M_{\Xi_{cc}}(1/2)&=&3.48(6)~,~~~~M_{\Xi_{bb}}(3/2)=9.94(91)~,
\label{eq:bagan}
\eea
and in \cite{BC}:
\bea
M_{\Xi_{bcu}}&=&6.86(28)~.
\label{eq:bc}
\eea
More recently \cite{ZHANG,WANG}, some results have been obtained using some particular choices of the interpolating currents. The predictions for $M_{\Xi^*_{cc}}$ and $M_{\Xi_{cc}}$ are in good agreement with the experimental candidate $M_{\Xi_{cc}}=3518.9$ \cite{PDG}. We shall also improve these previous predictions by working with the DRSR for estimating the mass ratio of the 3/2 over the 1/2 baryons and shall compare them with some potential model predictions \cite{RICHARD,BRAC,BC,VIJANDE}. 
\section{The interpolating currents and the two-point correlator}
\nin
\b {\bf For the spin 1/2  \boldmath$QQq$ baryons },  and following Ref. \cite{BAGAN}, we work with the lowest dimension currents:
\beq
J_{\Xi_Q}=\epsilon_{\alpha\beta\lambda}\left[(Q_\alpha^TC\gamma_5q_\beta)+b(Q_\alpha^TCq_\beta)\gamma_5\right]
Q_\lambda, \label{cur1}
\eeq
where $q\equiv d,s$ are light quark fields, $Q\equiv c,b$ are heavy quark fields, $b$ is {\it a priori} an arbitrary mixing parameter. Using the $b$-stability criterion of the QSSR results for the masses and couplings,  the optimal values of these observables have been found for:
\beq
b=-1/5~,
\label{eq:mixing1}
\eeq
in the case of light baryons \cite{JAMI2} and in the range \cite{BAGAN1,BAGAN2,BAGAN3,HBARYON}:
\beq
-0.5\leq b\leq 0.5~,
\label{eq:mixing2}
\eeq
for non-strange heavy baryons .
The corresponding two-point correlator reads:
\bea
S(q)&=&i\int d^4x~ e^{iqx}~ \la 0\vert {\cal T} \overline{J}_{\Xi_Q}(x)J_{\Xi_Q}(0)\vert 0\ra\nnb\\
&\equiv& \hat q F_1  +F_2~,
\label{eq:spin1/2}
\eea
where\,\footnote{We use the notation in the Landau and Lifchitz's book.}: $\hat q\equiv \slashed{q}$. The invariants $F_j~(j=1,2)$ obey the dispersion relation:
\beq
F_j(q^2)=\int_{(2M_Q+mq)^2}^\infty {dt\over t-q^2-i\epsilon}~{1\over\pi}{\rm Im} F_j(t)+\dots,
\eeq
where $\dots$ indicate subtraction constants and $-q^2\equiv Q^2>0$. 
\\
\b {\bf For the spin 3/2   \boldmath$QQq$ baryons}, we  also follow  Ref. \cite{BAGAN} and work with the interpolating currents:
\beq
J^\mu_{\Xi^*_Q}=
\sqrt{1\over3}\epsilon_{\alpha\beta\lambda}\big{[}
2(Q_\alpha^TC\gamma^\mu d_\beta)Q_\lambda+(Q_\alpha^TC\gamma^\mu Q_\beta)q_\lambda\big{]}
\label{cur}
\eeq
The corresponding two-point correlator reads:
\bea
S^{\mu\nu}(q)&=&i\int d^4x~ e^{iqx}~ \la 0\vert {\cal T} \overline{J}^\mu_{\Xi^*_Q}(x)J^\nu_{\Xi^*_Q}(0)\vert 0\ra\nnb\\
&\equiv& g^{\mu\nu}\ga \hat q F_1  +F_2\dr+\dots~, 
\label{eq:spin3/2}
\eea
\section{The  two-point correlator in QCD}
\nin
The expressions of the two-point correlator using the previous interpolating currents
have been obtained in the chiral limit $m_q=0$ and including the mixed condensate contributions by \cite{BAGAN}. In this paper, we extend
these results by including the linear strange quark mass corrections to the perturbative and $\la\bar ss\ra$ condensate contributions. We shall use the same
normalizations as in \cite{BAGAN}. \\
\b {\bf For the spin 1/2 baryons}, these corrections read:
\bea
{\rm Im} F_1^{m_s}\vert_{pert}&=&\frac{3m_s m_Q^3}{2^8 \:\pi^3} (1-b^2) \Big{[} 6  {\cal L}_v (2x^2-1)\nnb\\
 &&+ v \left( 6x+1+\frac{2}{x} \right) \Big{]}~,
\nnb\\
{\rm Im} F_2^{m_s}\vert_{pert}&=&\frac{m_s m_Q^4}{2^9 \:\pi^3} \Bigg{[} 12  {\cal L}_v \Big{[}2x(5b^2+2b+5) \nnb\\
&&-3(3b^2+2b+3)  \Big{]} 
+\nnb\\
&& v \Big{[} 12(5b^2+2b+5) \nnb\\
&&+\frac{4}{x}(5b^2+8b+5)
+ \frac{1}{x^2}(1-b)^2\Big{]}\Bigg{]}~ ,\nnb\\
{\rm Im} F_1^{m_s}\vert_{\bar ss}&=&{m_s\la \bar ss\ra\over 2^8\pi}\Bigg{[} v\Big{[} 7b^2+4b+7+\nnb\\
&&8\ga b^2+b+1\dr x\Big{]} +{3\over v}(1+b^2)\Bigg{]}~,\nnb\\
{\rm Im} F_2^{m_s}\vert_{\bar ss}&=&3{m_sm_Q\la \bar ss\ra\over 2^7\pi}\ga 1-b^2\dr \ga 3v+{1\over v}\dr~.
\label{eq:spin0.5}
\eea
\b {\bf For the spin 3/2 baryons}, these corrections read:
\bea
{\rm Im} F_1^{m_s}\vert_{pert}&=&\frac{m_s m_Q^3}{24 \:\pi^3 x} \Big{[} 6 \:{\cal L}_v \left(2 x^2-1\right) x + \nnb\\
&& v (6x^2+x+2) \Big{]}~,\nnb\\
{\rm Im} F_2^{m_s}\vert_{pert}&=&\frac{m_s m_Q^4}{192 \:\pi^3 x^2}\Big{[ }24 \:{\cal L}_v (x^2+6x-5)x^2+ \nnb\\
&&
v(12x^3+74x^2+10x+3) \Big{]}~,\nnb\\
{\rm Im} F_1^{m_s}\vert_{\bar ss}&=&-\frac{m_s \la\bar ss\ra}{96 \:\pi} \left[ v(4x-3) - \frac{5}{v} \right]~,\nnb\\
{\rm Im} F_2^{m_s}\vert_{\bar ss}&=&-\frac{m_s m_Q \la\bar ss\ra}{6 \:\pi v}(2x-1)~,
\label{eq:spin1.5}
\eea
with: 
\beq
x\equiv{m^2_Q\over t}~,~~v\equiv\sqrt{1-4x}~, ~~{\cal L}_v\equiv\log {\ga {1+v\over 1-v}\dr}~,
\eeq
\section{Form of the sum  rules and QCD inputs}
\label{sec:qssr}
\vspace*{-0.25cm}
 \nin
 We shall work with the exponential sum rules \cite{SVZ,BB,SNDR}:
 \beq
 {\cal F}_i{(\tau)}=\int_{t_q}^{\infty}dt~e^{-t\tau}~{1\over\pi}{\rm Im}F_{i}(t)~,~~~(i=1,2)~,
 \label{eq:laplace}
\eeq
($\tau\equiv 1/M^2$ is the sum rule variable) from which, one can derive the following ratios:
\bea
{\cal R}^q_i(\tau)\equiv-{d\over d\tau} \ln{{\cal F}_i}&=&{\int_{t_q}^{\infty}dt~t~
e^{-t\tau}~{\rm Im}F_{i}(t)\over \int_{t_q}^{\infty}dt~
e^{-t\tau}~{\rm Im}F_{i}(t)}~,~~~(i=1,2)~,\nnb\\
{\cal R}^q_{21}(\tau)\equiv {{\cal F}_2\over {\cal F}_1}&=&{\int_{t_q}^{\infty}dt~
e^{-t\tau}~{\rm Im}F_{2}(t)\over \int_{t_q}^{\infty}dt~
e^{-t\tau}~{\rm Im}F_{1}(t)}~,
\label{eq:ratio}
\eea
used in the sum rule literature for extracting the baryon masses. 
We parametrize the spectral function using the standard duality ansatz: ``one resonance"+ ``QCD continuum". The QCD continuum starts from a threshold $t_c$ and comes from the discontinuity of the QCD diagrams, which is consistent with a matching of the QCD and the experimental sides of the sum rules for large $t$. The value of $t_c$ is not arbitrary as its value obtained inside the region of $t_c$-stablity of the sum rule is correlated to the ground state mass and coupling \cite{fesr}. This simple duality model has been successfully tested in the literature when a complete data for the spectral functions are available, like e.g., $e^+e^-$ to I=1 hadrons or to charmonium data \cite{SNB}\,\footnote{More involved parametrizations of the continuum can also be proposed (see e.g. \cite{BIJNENS} for non-resonant final states within ChPT or \cite{LUCHA} for a  t-dependent $t_c$ model.). However,  it is easy to check in the harmonic oscillator model discussed in \cite{LUCHA} that the 5\% uncertainties induced e.g. by a $t$-dependent continuum model on the ratio of moments will be negligible in the double ratio of sum rule (DRSR) defined in Eq. (\ref{eq:2ratio13}) as the leading corrections coming from light-quark flavour-independent terms will largely cancel out. This cancellation of the QCD continuum contribution will be signaled by the large range of $t_c$-stability region obtained in our analysis. }.  Transferring the QCD continuum contribution to the QCD side of the sum rules, one obtains the finite energy inverse Laplace sum rules:
\bea
&&|\lambda_{B^{(*)}_q}|^2M_{B^{(*)}_q}~e^{-{M_{B^{(*)}_q}}^2\tau}=\int_{t_q}^{t_c}dt~
e^{-t\tau}~{1\over\pi}{\rm Im}F_{2}(t)~,
\nnb\\
&&|\lambda_{B^*_q}|^2~e^{-{M_{B^{(*)}_q}}^2\tau}=\int_{t_q}^{t_c}dt~
e^{-t\tau}~{1\over\pi}{\rm Im}F_{1}(t)~,
\lb{srm} 
\eea
where $\lambda_{B^{(*)}_q}$ and $M_{B^{(*)}_q}$ are the heavy baryon residue and mass from which one can derive the FESR analogue of the ratios of sum rules.  Consistently, we also take into account the SU(3) breaking at the  continuum threshold\,\footnote{As we have done an expansion in terms of $m_s$, the quark threshold has to be taken at $4m_Q^2$ but not at $(2m_Q+m_s)^2$.}:
 \bea
   \sqrt{t_c}\vert_{SU(3)}&\simeq&   \ga\sqrt{t_c}\vert_{SU(2)}\equiv \sqrt{t_c}\dr + \bar m_s~.
   \eea
$\bar m_{s}$  is the running strange quark mass. As we do an expansion in $m_s$, we take the threshold $t_q=4m_Q^2$ for consistency. $m_Q$ is the heavy quark mass, which we shall take in the range covered by the running and on-shell mass (see Table \ref{tab:param}) because of its ambiguous definition  when working to LO.  
At the $\tau$-stability point of the FESR analogue of the ratios of sum rules, one obtains:
\beq
M_{B^{(*)}_q}\simeq \sqrt{{\cal R}^q_i}\simeq {\cal R}^q_{21}~,~~~~~(i=1,2)~.
\eeq
These predictions lead to a  typical uncertainty of 10-15\% \cite{BAGAN,BAGAN2,BAGAN3}, which are not competitive compared with predictions from some other approaches, especially from potential models \cite{RICHARD}. 
In order to improve the QSSR predictions, we work with the double ratios of finite energy sum rules (DRSR)\,\footnote{Analogous DRSR quantities have been used successfully (for the first time) in \cite{SNhl} for studying the mass ratio of the $0^{++}/0^{-+}$ and $1^{++}/1^{--}$ B-mesons, in \cite{SNFBS} for extracting  $f_{B_s}/f_B$, in \cite{SNFORM} for estimating the $D\to K/D\to \pi$ semi-leptonic form factors and in \cite{SNme+e-} for extracting the strange quark mass from the $e^+e^-\rar I=1,0$ data.}:
\beq
r^{sd}_i\equiv \sqrt{{\cal R}^s_i\over {\cal R}^d_i}~~~(i=1,2)~~;~~~~~~~~~~r^{sd}_{21}\equiv {{\cal R}^s_{21}\over {\cal R}^d_{21}}~.
\label{eq:2ratio}
\eeq
which take directly into account the SU(3) breaking effects. These quantities are obviously less sensitive to the choice of the heavy quark masses and to the value of the continuum threshold than the simple ratios ${\cal R}_i$ and  ${\cal R}_{21}$\,\footnote{One may also work with the double ratio of moments ${\cal M}_n$ based on different derivatives  at $q^2=0$ \cite{SNhl}. However, in this case the OPE is expressed as an expansion in $1/m_Q$, which for a LO expression of the QCD correlator is more affected by the definition of the heavy quark mass to be used.}. For the numerical analysis we shall  introduce the RGI quantities $\hat\mu$ and $\hat m_q$ \cite{FNR}:
\bea
\bar m_q(\tau)&=&{\hat m_q\over \ga -\log{ \sqrt{\tau}\Lambda}\dr^{2/{-\beta_1}}}\nnb\\
{\la\bar qq\ra}(\tau)&=&-{\hat \mu_q^3 \ga-\log{ \sqrt{\tau}\Lambda}\dr^{2/{-\beta_1}}}\nnb\\
{\la\bar qGq\ra}(\tau)&=&-{\hat \mu_q^3 \ga-\log{ \sqrt{\tau}\Lambda}\dr^{1/{-3\beta_1}}}M_0^2~,
\eea
where $\beta_1=-(1/2)(11-2n/3)$ is the first coefficient of the $\beta$ function for $n$ flavours. We have used the quark mass and condensate anomalous dimensions reviewed in \cite{SNB}.  We shall use the QCD parameters in Table \ref{tab:param}:\\
-- We shall not include the $1/q^2$ term discussed in \cite{CNZ,ZAK},which is consistent with the LO approximation used here as the latter has been motivated for a phenomenological parametrization  of the larger order terms of the QCD series.  \\
  -- We have used the value of $\kappa\equiv\la\bar ss\ra/\la\bar dd\ra$ from \cite{HBARYON}  which we consider as improvements of the ones from light meson systems \cite{SNmass,SNB,SNSU3,DOM}, where the one from the scalar channel suffers from the unknown nature of the $\kappa$ meson, while the one from the pseudoscalar channel depends on the theoretical appreciation of the $\pi'$ meson contribution into the spectral function \cite{BIJNENS,SNB}. However, the different estimates agree each others within the errors. To be conservative, we have multiplied the original error in \cite{HBARYON} by 2.\\
 -- For the gluon condensate, we have used the estimate from heavy quarkonia  and $e^+e^-$ data \cite{SNTAU,LNT,SNI,fesr,YNDU,SNHeavy,BB,SNH10}. We do not expect that the estimate from $\tau$-decays is reliable as its contribution acquires an extra-$\alpha_s$ term in the $\tau$ width compared to the one in the two-point correlator \cite{BNP}, while its value, in this process, can also be affected by the treatments of the large order PT series \cite{SNTAU,CNZ}. However, the effect of the gluon condensate is not important in our analysis of the SU(3) breaking as it disappears like some other flavour-independent contributions in the DRSR.\\
 -- For the heavy quark masses, we use the range spanned
 by the running $\overline{MS}$ mass $\overline{m}_Q(M_Q)$  and the on-shell mass from QSSR compiled in page 602,  603 of the book in \cite{SNB}. \\


{\scriptsize
\begin{table}[hbt]
\setlength{\tabcolsep}{0.5pc}
 \caption{\scriptsize    QCD input parameters. The values of $\Lambda$, $\hat m_s$ and $\mu_d$ have been obtained from $\alpha_s(M_\tau)=0.325(8)$ \cite{SNTAU} and from the running masses: $\overline{m}_s(2)=96.1(4.8)$ MeV and $\overline{m}_d(2)=5.1(2)$ MeV \cite{SNmass}. The original errors for $\kappa$ and $\la\alpha_s G^2\ra$ have been multiplied by 2.
    }
    {\small
\begin{tabular}{lll}
&\\
\hline
Parameters&Values& Ref.    \\
\hline
$\Lambda(n_f=4)$& $(324\pm 15)$ MeV &\cite{SNTAU,PDG}\\
$\Lambda(n_f=5)$& $(194\pm 10)$ MeV &\cite{SNTAU,PDG}\\
$\hat m_s$&$(114.5\pm 20.8)$ MeV&\cite{SNme+e-,SNmass,SNB,PDG}\\
$\hat \mu_d$&$(263\pm 7)$ MeV&\cite{SNmass,SNB}\\
$\kappa\equiv{\la\bar ss\ra}/{\la\bar dd\ra}$&$(0.74\pm 0.06)$&\cite{SNmass,SNB,HBARYON,DOM} \\
$M_0^2$&$(0.8 \pm 0.1)$ GeV$^2$&\cite{JAMI2,HEID,SNhl}\\
$\la\alpha_s G^2\ra$& $(6\pm 2)\times 10^{-2}$ GeV$^4$&\cite{SNTAU,LNT,SNI,fesr,YNDU,SNHeavy,BB,SNH10}\\
$m_c$&$(1.26\sim1.47)$ GeV &\cite{SNB,SNmass,SNHmass,PDG,SNH10}\\
$m_b$&$(4.22\sim4.72)$ GeV &\cite{SNB,SNmass,SNHmass,PDG,SNH10}\\
\hline
\end{tabular}
}
\label{tab:param}
\end{table}
}
\vspace*{-0.5cm}
\nin
\section{The  \boldmath $\Xi^*_{QQ}/\Xi_{QQ }$ mass ratio}
\vspace*{-0.25cm}
\begin{figure}[hbt]
\begin{center}
\includegraphics[width=5.2cm]{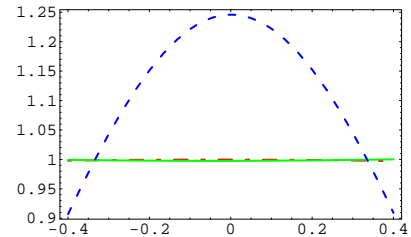}
\vspace*{-0.3cm}
\caption{\footnotesize  {\bf Charm quark:} $b$-behaviour of the different DRSR given $\tau=0.8$ GeV$^{-2}$ and $t_c=25$ GeV$^2$.  $r^{3/1}_1$ dot-dashed line (red); $r^{3/1}_2$ continuous line (green); $r^{3/1}_{12}$ dashed line (blue). 
We have used $m_c=1.26$ GeV and the other QCD parameters in Table~\ref{tab:param}. }
\label{fig:r13b}
\end{center}
\end{figure} 
  \nin
 We extract the mass ratio using the DRSR analogue of the one in Eq. (\ref{eq:2ratio}) which we denote by:
 \beq
r^{3/1}_i\equiv \sqrt{{\cal R}^3_i\over {\cal R}^1_i}~:~~i=1,2~~;~~~~~~~~r^{3/1}_{21}\equiv {{\cal R}^3_{21}\over {\cal R}^1_{21}}~,
\label{eq:2ratio13}
\eeq
where the upper indices 3 and 1 correspond respectively to the spin 3/2 and 1/2 channels. We use the QCD expressions of the two-point correlators given by \cite{BAGAN} which we have checked. We notice like \cite{BAGAN} that the mixed quark condensate contribution has a term which behaves like $1/v^3$ (where $v$ is the heavy quark velocity), which signals a coulombic correction and would require a complete treatment of the non-relativistic coulombic corrections which is beyond the aim of this paper.  Therefore, in our analysis, we truncate the QCD series  at the dimension-4 condensates until which we have calculated the $m_s$ corrections. We shall only include the effect of the mixed condensate (if necessary) for controlling the accuracy of the approach or for improving the $\tau$ or/and $t_c$ stability of the analysis. 
  \subsection*{\b The charm quark channel to lowest order in $\alpha_s$}
\begin{figure}[hbt]
\begin{center}
\includegraphics[width=5.2cm]{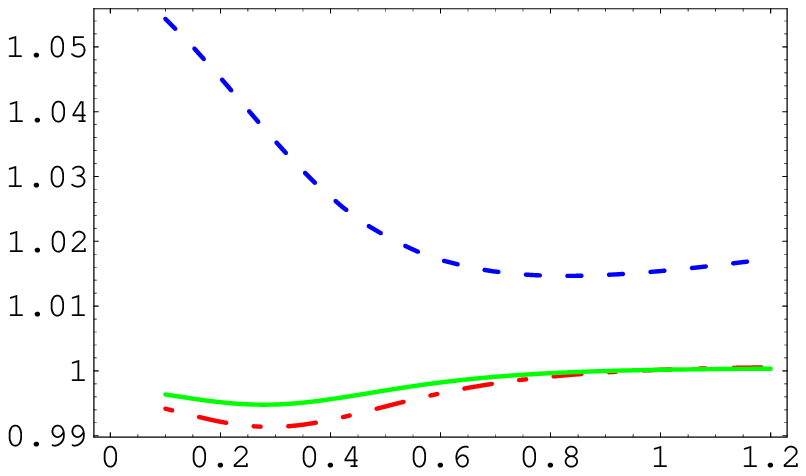}
\includegraphics[width=5.2cm]{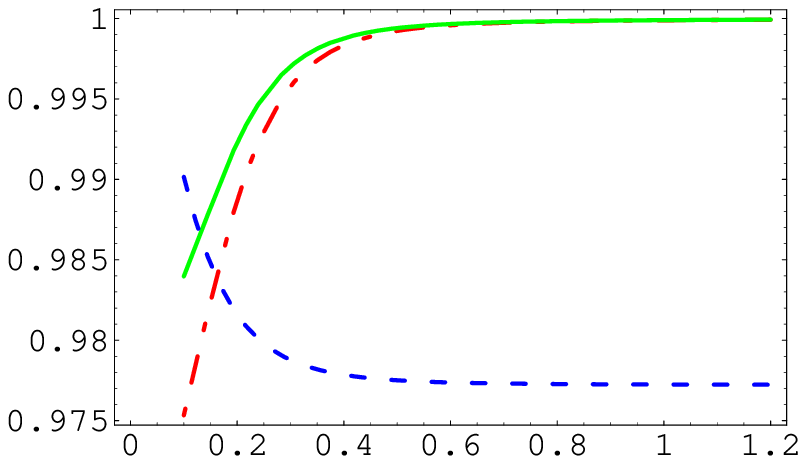}
\vspace*{-0.3cm}
\caption{\footnotesize {\bf Charm quark:}
a) $\tau$-behaviour of $r^{3/1}_1$: dot-dashed line (red), $r^{3/1}_2$ : continuous line (green) and $r^{3/1}_{12}$ dashed line (blue) with $b=-0.35$ and $t_c=25$ GeV$^{2}$. b) the same as a) but when the mixed condensate is included.}
\label{fig:r13tau}
\end{center}
\vspace*{-.5cm}
\end{figure}

\begin{figure}[hbt]
\begin{center}
\includegraphics[width=5.2cm]{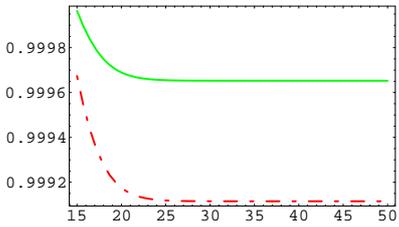}
\vspace*{-0.3cm}
\caption{\footnotesize {\bf Charm quark:}
$t_c$-behaviour of $r^{3/1}_1$: dot-dashed line (red) and $r^{3/1}_2$ : continuous line (green) with $b=-0.35$ and $\tau=0.8$ GeV$^{-2}$.
}
\label{fig:r13tc}
\end{center}
\vspace*{-.5cm}
\end{figure}

\nin
Fixing $\tau=$ 0.8 GeV$^{-2}$ and $t_c=25$ GeV$^2$, which are inside the $\tau$- and $t_c$-stability regions (see Figs. \ref{fig:r13tau} and  \ref{fig:r13tc}), we show in Fig. \ref{fig:r13b} the $b$-behaviour of $r^{3/1}$ which shows that $r^{3/1}_1$ and $r^{3/1}_2$ are very stable but not $r^{3/1}_{12}$. We then disfavour $r^{3/1}_{12}$. Some common solutions are obtained for:
\beq
b\simeq -0.35~,~~~~~~~{\rm and}~~~~~~~b\simeq +0.2~,
\eeq
which are inside the range given in Eq. (\ref{eq:mixing2}). For definiteness, we fix $b=-0.35$ (the other value $b=0.2$ gives the same result) and study the  $\tau$-dependence of the result in Fig. \ref{fig:r13tau}. We have checked in Fig. \ref{fig:r13tau}b that the inclusion of the mixed condensate contribution does not affect 
the result from $r^{3/1}_i~(i=1,2)$ obtained  by retaining only the dimension-4 condensates (Fig. \ref{fig:r13tau}a) but affects the one from $r^{3/1}_{12}$. Therefore, we shall only retain the results from $r^{3/1}_i~(i=1,2)$ and show their
$t_c$-dependence  in Fig. \ref{fig:r13tc}. 
\nin
   \vspace*{-.5cm}
\nin
 The large stability in $t_c$ confirms our expectation of the weak $t_c$-dependence of the DRSR and then on the non-sensitivity of the results on the exact form of the QCD continuum including an eventual slight $t$-dependence of $t_c$ advocated in \cite{LUCHA}. In these figures, we have used $m_c=1.26$ GeV. We have also checked that the results are insensitve to the change of the charm mass to $m_c=1.47$ GeV. 
From these previous analysis, we deduce to lowest order from $r^{3/1}_i~(i=1,2)$:
\beq
{M_{\Xi^*_{cc}}\over M_{\Xi_{cc }}}= 0.9994(3)~.
\label{eq:chicc}
\eeq
The tiny error is the quadratic sum due to $\la\alpha_s G^2\ra$, $m_c$ and $\alpha_s$. 
  \subsection*{\b The bottom quark channel to lowest order in $\alpha_s$}
\nin
\begin{figure}[hbt]
\begin{center}
\includegraphics[width=5.2cm]{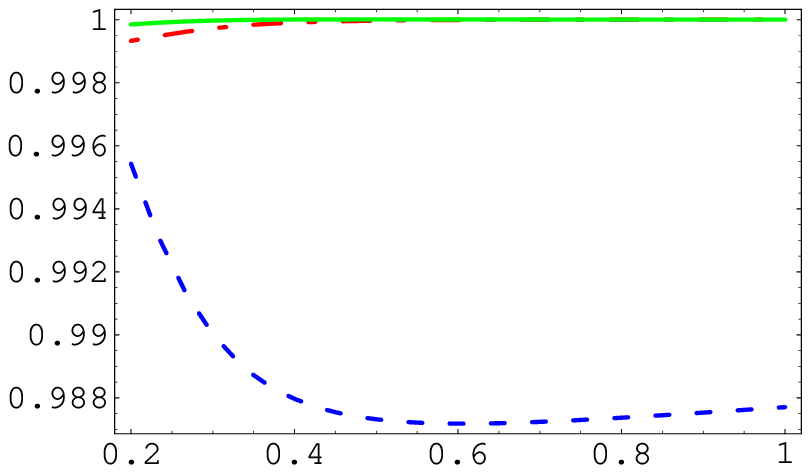}
\includegraphics[width=5.2cm]{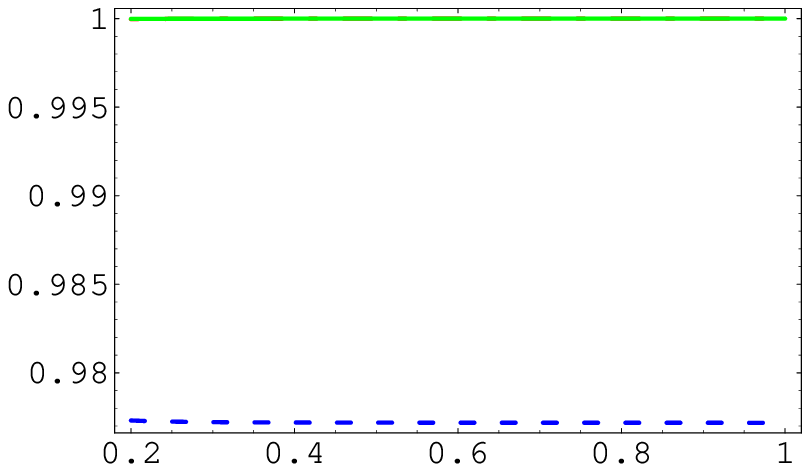}
\vspace*{-0.3cm}
\caption{\footnotesize {\bf Bottom quark:}
a) $\tau$-behaviour of $r^{3/1}_1$: dot-dashed line (red), $r^{3/1}_2$: continuous line (green)and  $r^{3/1}_{12}$: dashed line (blue)  with $b=-0.35$ and $t_c=100$ GeV$^{2}$; b) the same as a) but when the mixed condensate is included into the OPE.}
\label{fig:r13btau}
\end{center}
\vspace*{-.5cm}
\end{figure} 
\nin
\vspace*{-.5cm}
\begin{figure}[hbt]
\begin{center}
\includegraphics[width=5.2cm]{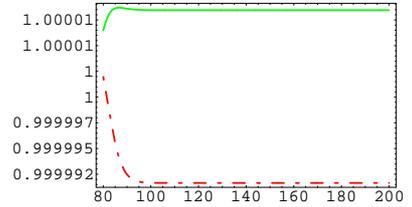}
\vspace*{-0.3cm}
\caption{\footnotesize {\bf Bottom quark:}
$t_c$-behaviour of $r^{3/1}_1$: dot-dashed line (red) and $r^{3/1}_2$ : continuous line (green) with $b=-0.35$ and $\tau=0.6$ GeV$^{-2}$.}
\label{fig:r13btc}
\end{center}
\vspace*{-0.5cm}
\end{figure} 
\nin
We extend the analysis to the case of the bottom quark. The corresponding curves are qualitatively similar to the charm quark one.  
We take $b=-0.35$ like in the case of the charm quark. 
The $\tau$-stability is reached for $\tau\geq 0.6$ GeV$^{-2}$ as shown in Fig. \ref{fig:r13btau}, where we also see that $r^{3/1}_{12}$ is more affected by the mixed condensate contributions than $r^{3/1}_i$. Therefore, we shall eliminate it from our choice. Another argument raised later about the radiative corrections does not also favour $r^{3/1}_{12}$.  In Fig. \ref{fig:r13btau}, we study  the $t_c$-stability of $r^{3/1}_i$ which is reached for $t_c\geq 95$ GeV$^2$.
Within these optimal conditions, one deduces from $r^{3/1}_i$ to lowest order:
\beq
{M_{\Xi^*_{bb }}\over M_{\Xi_{bb}}}= 1.0000~.
\label{eq:chibb}
\eeq
  \subsection*{\b Estimate of the ${\cal O}(\alpha_s)$ corrections}
\nin
Radiative corrections due to $\alpha_s$ are known to be large in the baryon two-point correlators \cite{JAMI2,KORNER}. However, one can easily inspect that in the simple ratios ${\cal R}^3_i$ and ${\cal R}^1_i$ these huge corrections cancel out, while its only remain the one induced by the anomalous dimension of
the baryon operators. Including the anomalous dimension $\gamma$= 2(resp -2/3) for the spin 1/2 (resp 3/2) baryons \cite{JAMI2}, one can generically write the  PT expressions of the moment sum rule defined in Eq. (\ref{eq:laplace}) to leading order in $t/m_Q^2$, which is a crude approximation but very informative: 
\beq
{\cal F}_i(\tau)|_{pert}\approx \ga\alpha_s(\tau)\dr^{-{\gamma\over \beta_1} }A_i~\tau^{-3}\ga 1+K_i{\alpha_s\over\pi}\dr,
\label{eq:nlo}
\eeq
where $\beta_1$ is the first coefficient of the $\beta$-function; $A_i$ is a known LO expression; $K_i$ is the radiative correction which is known in some cases of light and heavy baryons  \cite{JAMI2,KORNER}. From the previous expression in Eq. (\ref{eq:nlo}), one can derive the ratio of sum rules defined in Eq. (\ref{eq:ratio}) and then the DRSR in Eq. (\ref{eq:2ratio}):
\beq
r_i^{3/1}|^{NLO}_{pert}\simeq  r_i^{3/1}|^{LO}_{pert}\times \Bigg{[}1+{2\over 9}{\alpha_s\over\pi}+{\cal O}\ga\alpha_s^2, M^2_Q\tau\dr\Bigg{]}~.
\eeq
\vspace*{-0.5cm}
\begin{figure}[hbt]
\begin{center}
\includegraphics[width=5.2cm]{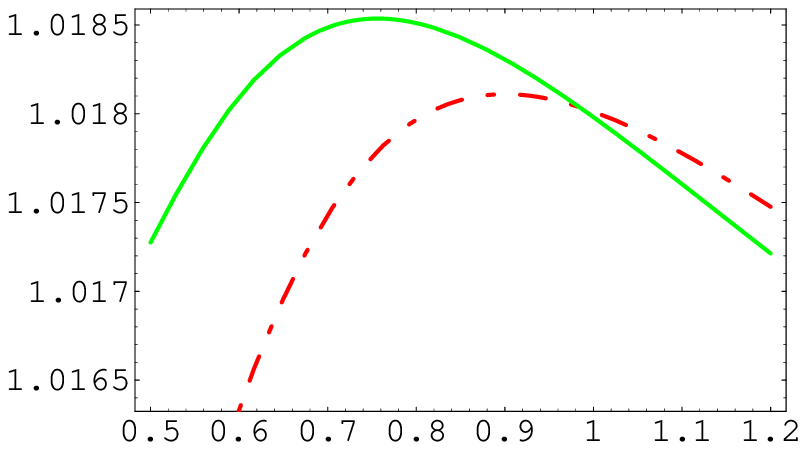}
\includegraphics[width=5.2cm]{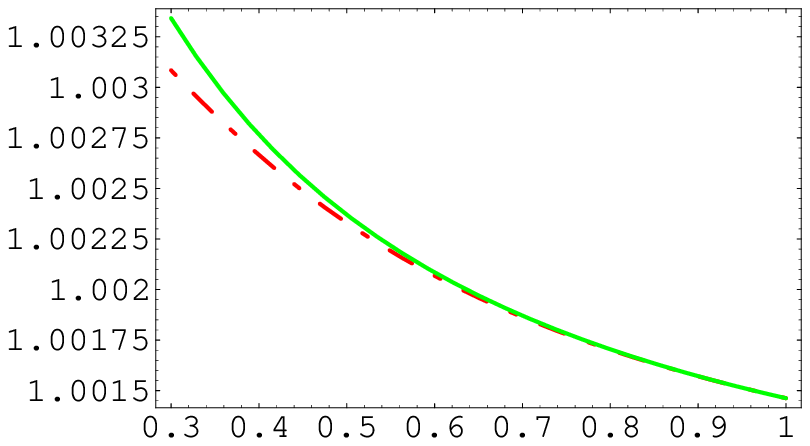}
\vspace*{-0.3cm}
\caption{\footnotesize {\bf Charm quark:}
a) $\tau$-behaviour of $r^{3/1}_1$: dot-dashed line (red) and $r^{3/1}_2$ : continuous line (green) with $b=-0.35$, $t_c= 25$ GeV$^2$ where radiative corrections have been included. {\bf Bottom quark:}  b) the same as in a) but for the bottom quark. We use $b=-0.35$ and  $t_c= 100$ GeV$^2$}
\label{fig:alphas}
\end{center}
\vspace*{-0.5cm}
\end{figure} 
\nin
It is important to notice for $r_i^{3/1}$ that the radiative correction has been only induced by the ones due to the anomalous dimensions, while the one due to $K_i$ cancels out to this order. This is not the case of $r_{12}^{3/1}$ where the radiative correction is only due to $K_2-K_1$ and needs to be evaluated which is beyond the aim of this letter.  Therefore, in the following, we shall only consider the results from $r_i^{3/1}$. In our numerical analysis, we shall include the $\alpha_s$ correction into the complete LO expressions of the correlators. We show the $\tau$-dependence of the DRSR in Fig. \ref{fig:alphas}. We shall take the range
of $\tau$-values where the LO expressions have $\tau$-stability which is (0.7-1) GeV$^{-2}$ for charm and (0.5-0.8) GeV$^{-2}$ for bottom (see Figs. \ref{fig:r13tau} and \ref{fig:r13btau}). One can also notice that the NLO DRSR for charm presents a  $\tau$-extremum in the above range (0.7-1) GeV$^{-2}$ of $\tau$ rendering its prediction more reliable than for the bottom channel case. We can deduce :
\beq
{M_{\Xi^*_{cc}}\over M_{\Xi_{cc }}}= 1.0167(10)_{\alpha_s}(16)_{m_c},~{M_{\Xi^*_{bb }}\over M_{\Xi_{bb}}}=1.0019(3)_{\alpha_s}(2)_{m_b}.
\label{eq:chicb}
\eeq
This would correspond to the mass-splittings (in units of MeV):
\beq
{M_{\Xi^*_{cc }}- M_{\Xi_{cc}}}= 59(7)~,~~~~~~~~~{M_{\Xi^*_{bb }}- M_{\Xi_{bb}}}= 19(3)~,
\label{eq:chibcmass}
\eeq
if one uses the experimental value 3.52 GeV of the  $\Xi_{cc}$ mass which agrees with the QSSR prediction in Eq. (\ref{eq:bagan}). For the $\Xi_{bb}$ mass, we have used the central value 9.94 GeV in  Eq. (\ref{eq:bagan}) . 
The $ccq$ mass-splitting  is comparable with the one of about 70 MeV from potential models \cite{BC,RICHARD} but larger than the one of about  24 MeV obtained in \cite{VIJANDE} . The $bbq$ mass-splitting also agrees with potential models  and seems to indicate a $1/M_b$ behaviour which is also seen on the lattice \cite{LATT}. Our result excludes the possibility that $M_{\Xi^*_{QQ}}\geq M_{\Xi_{QQ}}+m_\pi$, indicating that it can only decay
electromagnetically:
\beq
M_{\Xi^*_{QQ }}\to M_{\Xi_{QQ }}\gamma~,~~~~~~~~~~~~~~~~~M_{\Xi^*_{QQ}}\not\to M_{\Xi_{QQ }} \pi~.
\eeq
A future discovery of the ${\Xi^*_{cc }}$ and ${\Xi^*_{bb }}$ can infirm or support our predictions given to
that order of QCD perturbative series. 
We consider the previous results as an improvement of the former ones deduced from the mass values in Eq. (\ref{eq:bagan}) obtained by \cite{BAGAN}:
\beq
{M_{\Xi^*_{c c}}\over M_{\Xi_{cc }}}\simeq 1.03\pm 0.03~,~~~~~~~~~{M_{\Xi^*_{bb }}\over M_{\Xi_{bb }}}= 1.04\pm 0.23~.
\eeq
\section{The  \boldmath $\Omega_{QQ}/\Xi_{QQ }$ mass ratio}
\nin
\begin{figure}[hbt]
\begin{center}
\includegraphics[width=5.2cm]{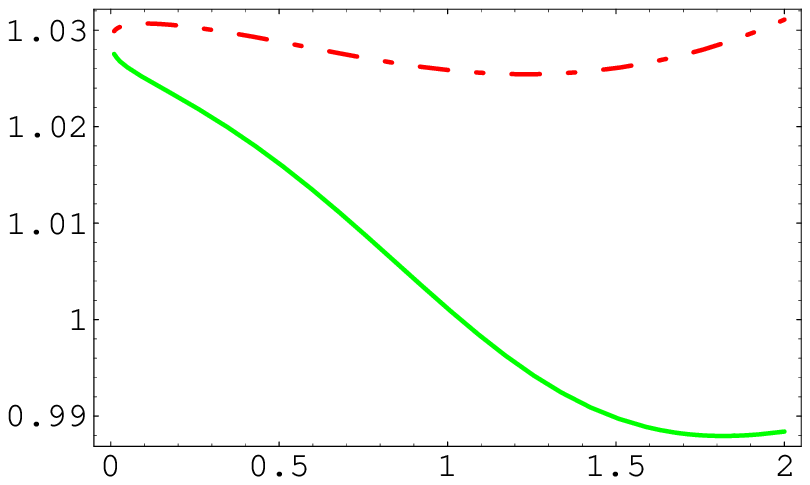}
\includegraphics[width=5.2cm]{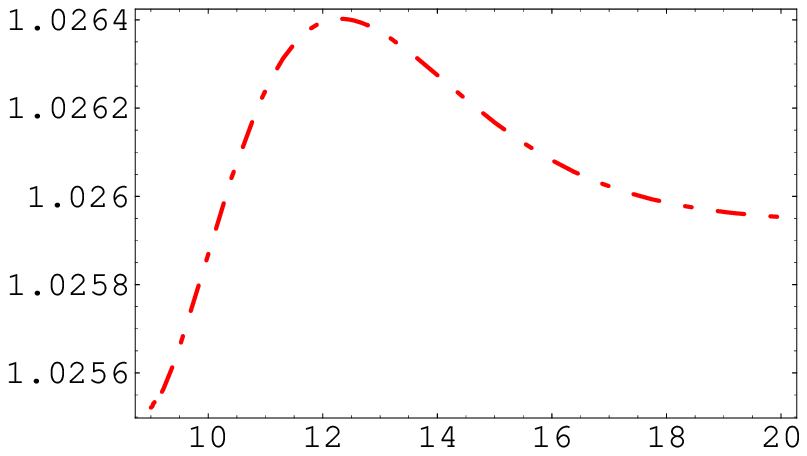}
\vspace*{-0.3cm}
\caption{\footnotesize {\boldmath$\Omega_{cc}/\Xi_{cc}:$} a) $\tau$-behaviour of $r^{sd}_2(cc)$: continuous line (green) and $r^{sd}_1(cc)$: dot-dashed line (red) in the charm quark channel for $b=-0.35$, $t_c=12$ GeV$^{2}$ and $m_c=1.26$ GeV. 
b)  $t_c$-behaviour of $r^{sd}_1(cc)$ for $\tau=1$ GeV$^{-2}$: dot-dashed line (red) }
\label{fig:chic_tau}
\end{center}
\end{figure} 
\nin
  \vspace*{-0.5cm}
\begin{figure}[hbt]
\begin{center}
\includegraphics[width=5.2cm]{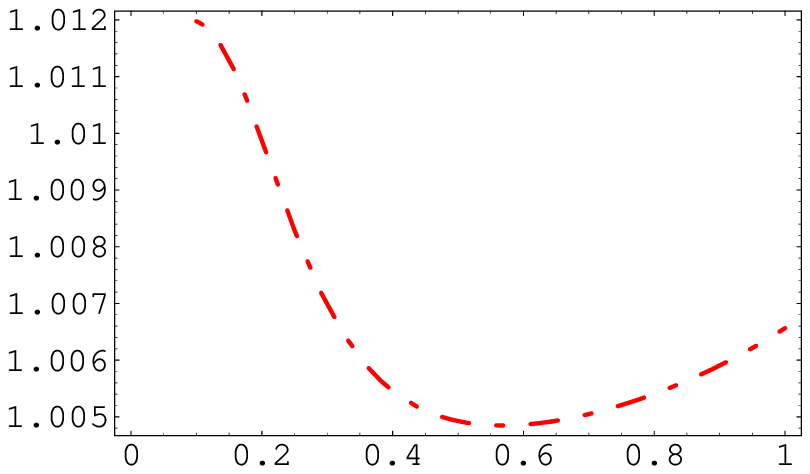}
\includegraphics[width=5.2cm]{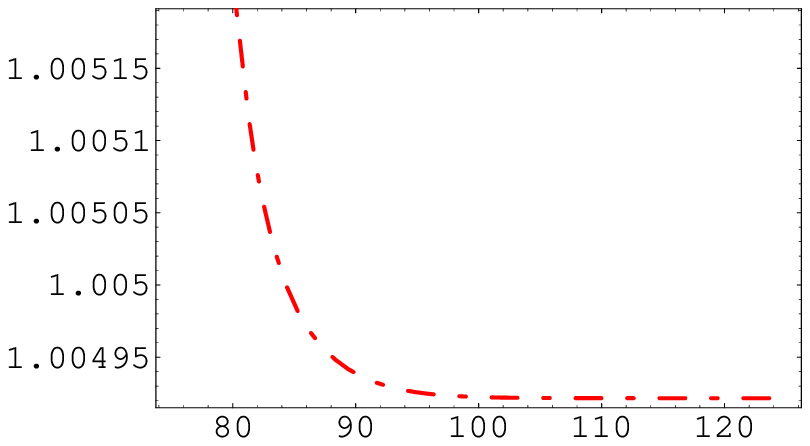}
\vspace*{-0.3cm}
\caption{\footnotesize {\boldmath$\Omega_{bb}/\Xi_{bb}:$} a) $\tau$-behaviour of $r^{sd}_1(bb)$: dot-dashed line (red) in the bottom quark channel for $b=-0.35$, $t_c=100$ GeV$^{2}$ and $m_b=4.22$ GeV. 
b)  $t_c$-behaviour of $r^{sd}_1(bb)$ for $\tau=0.5$ GeV$^{-2}$: dot-dashed line (green) }
\label{fig:chib_tau}
\end{center}
\vspace*{-.5cm}
\end{figure} 
\nin
 We use the DRSR in Eq. (\ref{eq:2ratio}) where their QCD expressions can be obtained from the one of the two-point correlator in \cite{BAGAN} and the new quark mass corrections in Eq. (\ref{eq:spin0.5}). One can also deduce from Eq. (\ref{eq:nlo}) that the light-flavour independent radiative corrections including the one due to the anomalous dimensions disappear in the SU(3) breaking DRSR, while the most relevant radiative corrections are the one corresponding to the $m_s$ and $\la \bar ss\ra$ terms which are beyond the scope of the LO analysis in this paper.  We show in Fig. \ref{fig:chic_tau}a the 
$\tau$-behaviour of the DRSR for $m_c=1.26$ GeV and $b=-0.35$ for a given $t_c=10$ GeV$^2$. We have not shown $r_{12}^{sd}(cc)$ which is the lesser stable among the three. We see that the most stable result is given by $r_1^{sd}(cc)$. We show in Fig. \ref{fig:chic_tau}b the $t_c$-behaviour of $r_1^{sd}(cc)$ for a given $\tau=1$ GeV$^{-2}$. 
 We deduce from the previous analysis:
 \beq
 r^{sd}_{1}(cc)\equiv {M_{\Omega_{cc}}\over M_{\Xi_{cc}}}=1.026(5)_{m_c}(2)_{\bar ss}(4)_{m_s}~,
 \eeq
 where the sub-indices indicate the different sources of errors (the parameters not mentioned induce negligible errors). This ratio corresponds to :
 \beq
 M_{\Omega_{cc}}- M_{\Xi_{cc}}= 92(24) ~{\rm MeV}~,
 \label{eq:omegac}
 \eeq
 where we have taken the experimental value $M_{\Xi_{cc}}\simeq 3.52$ GeV from  \cite{PDG}.
 The errors induced by the other parameters in Table \ref{tab:param} are negligible. 
We perform an analogous analysis in the $b$-channel, which we show in Fig. \ref{fig:chib_tau}. In this case, we obtain:
 \beq
 r^{sd}(bb)\simeq 1.0049(7)_{m_b}(3)_{\bar ss}(10)_{m_s}~,
 \eeq
 which corresponds to:
 \beq
 M_{\Omega_{bb}}- M_{\Xi_{bb}}=49(13)~{\rm MeV}~,
 \label{eq:omegab}
 \eeq
 when we take the value $M_{\Xi_{bb}}\simeq$ 9.94 GeV from \cite{BAGAN}.
 Our results indicate an approximate decrease like $1/m_Q$ of the mass splittings from the $c$ to the $b$ quark channels. This behaviour can be qualitatively understood from the QCD expressions of the corresponding correlator, where the $m_s$ corrections enter like $m_s/m_Q$, and which  can be checked  using some alternative methods.
\begin{figure}[hbt]
\begin{center}
\includegraphics[width=5.2cm]{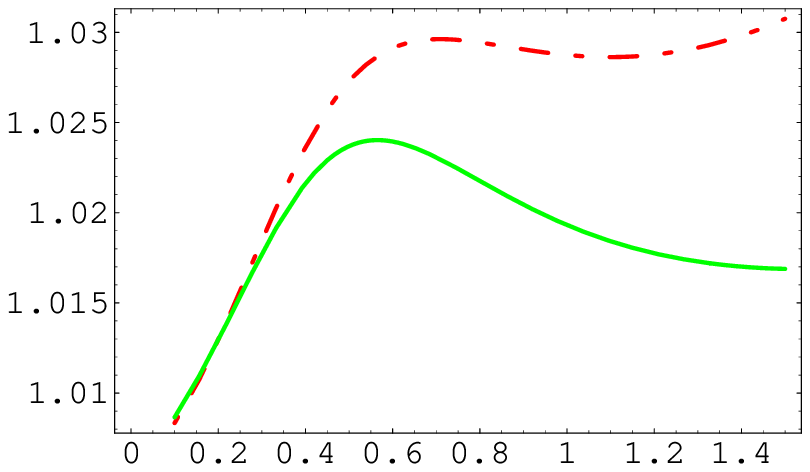}
\includegraphics[width=5.2cm]{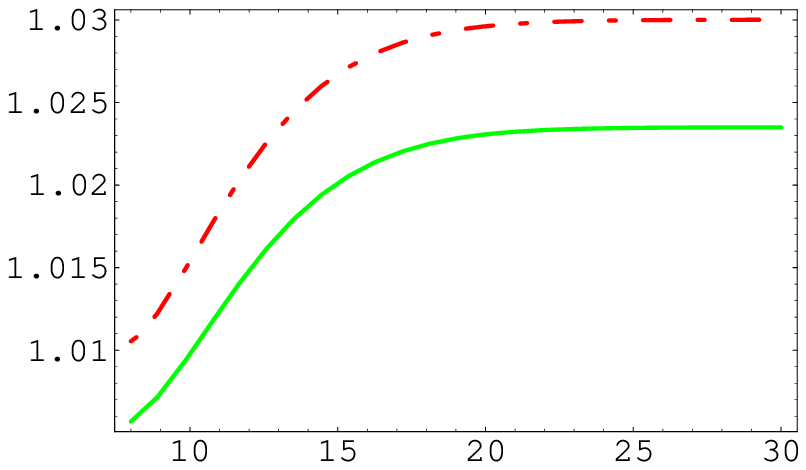}
\vspace*{-0.3cm}
\caption{\footnotesize  {\boldmath$\Omega^*_{cc}/\Xi^*_{cc}:$} a) $\tau$-behaviour of 
$r^{sd}_1(cc)^*$: dot-dashed line (red) and $r^{sd}_2(cc)^*$: continuous line (green)  in the charm quark channel for
 $t_c=20$ GeV$^{2}$ and $m_c=1.26$ GeV. 
b)  $t_c$-behaviour of  $r^{sd}_1(cc)^*$ and  $r^{sd}_2(cc)^*$for $\tau=0.7$ GeV$^{-2}$ }
\label{fig:chistc}
\end{center}
\vspace*{-.5cm}
\end{figure} 
\nin
\begin{figure}[hbt]
\begin{center}
\includegraphics[width=5.2cm]{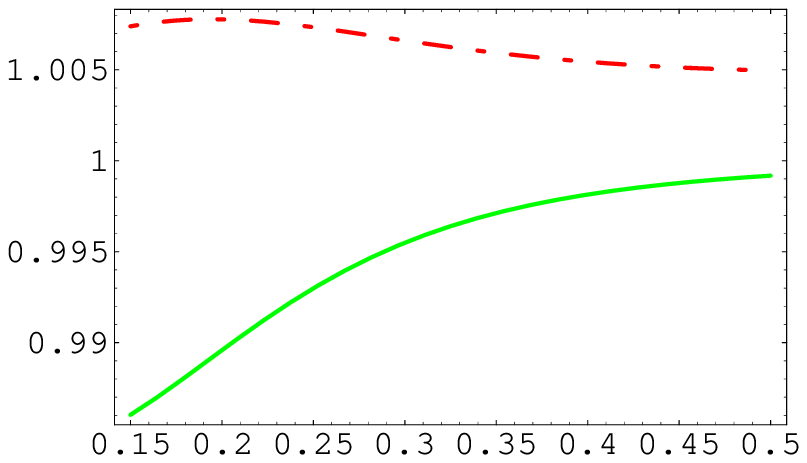}
\includegraphics[width=5.2cm]{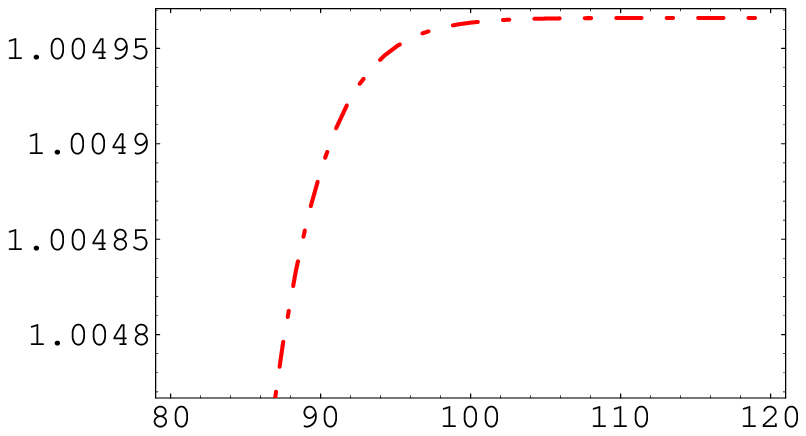}
\vspace*{-0.3cm}
\caption{\footnotesize  {\boldmath$\Omega^*_{bb}/\Xi^*_{bb}:$} a) $\tau$-behaviour of 
 $r^{sd}_1(bb)^*$: dot-dashed line (red) and of $r^{sd}_2(bc)$: continuous line (green) in the bottom quark channel for  $t_c=100$ GeV$^{2}$ and $m_b=4.22$ GeV. 
b) $t_c$-behaviour of  $r^{sd}_1(bb)^*$ for $\tau=0.5$ GeV$^{-2}$ }
\label{fig:chistb}
\end{center}
\vspace*{-.5cm}
\end{figure} 
\nin
\section{The  \boldmath $\Omega^*_{QQ}/\Xi^*_{QQ}$ mass ratio}
\nin
We pursue our analysis for the spin 3/2 baryons. The QCD expression of the ratios of moments can be obtained from the ones of the two-point correlator in \cite{BAGAN} and the new mass corrections given in Eq. (\ref{eq:spin1.5}). Including the contributions of the dimension-4 condensates, we show your analysis in Fig. \ref{fig:chistc}. One can see in Fig. \ref{fig:chistc}a that $r^{sd}_1$ and  $r^{sd}_2$ are quite stable versus $\tau$ from $\tau\geq 0.4$ GeV$^{-2}$.  In Fig. \ref{fig:chistc}b, we show the $t_c$-behaviour of  $r^{sd}_1$ and $r^{sd}_2$ given $\tau$. We deduce at the stability regions:
 \beq
 r^{sd}(cc)^*\equiv {M_{\Omega^*_{cc}}\over M_{\Xi^*_{cc}}}=1.026(4)_{\bar ss}(4)_{m_s}(6)_{m_c}(1)_{t_c}~,
 \eeq
where the errors coming from other parameters than  ${\bar ss}$ are negligible. This implies:
 \beq
 M_{\Omega^*_{cc}}- M_{\Xi^*_{cc}}= 94(27) ~{\rm MeV}~,
\label{eq:omega*c}
 \eeq
where we have used $M_{\Xi^*_{cc}}\simeq 3.58$ GeV from Eq. (\ref{eq:omegac}) and the experimental value of $M_{\Xi_{cc}}$. We show in Fig. \ref{fig:chistb} the analogous analysis for the bottom channel. We deduce:
 \beq
 r^{sd}(bb)^*\equiv {M_{\Omega^*_{bb}}\over M_{\Xi^*_{bb}}}=1.0050(3)_{\bar ss}(10)_{m_s}(4)_{\tau}(10)_{m_b}~,
 \eeq
where the error is again mainly due to ${\la\bar ss\ra}$, the others being negligible. This implies:
 \beq
 M_{\Omega^*_{bb}}- M_{\Xi^*_{bb}}= 50(15) ~{\rm MeV}~,
 \label{eq:omega*b}
 \eeq
where we have used $M_{\Xi^*_{bb}}\simeq 9.96$ GeV using our prediction in the previous section. 
This result agrees with the potential model one of about 60 MeV given in \cite{BC}. Again like in the case of spin 1/2 baryons, the SU(3) mass-differences appears to behave like  the inverse of the heavy quark masses, which can be inspected from the QCD expressions of the two-point correlator. 
One can also observe that the mass-splittings are almost the same for the spin 1/2 and spin 3/2 baryons.
\section{The  \boldmath $\Omega_{bc}/\Xi_{bc}$ mass ratio}
\nin
  \vspace*{-0.5cm}
\begin{figure}[hbt]
\begin{center}
\includegraphics[width=5.2cm]{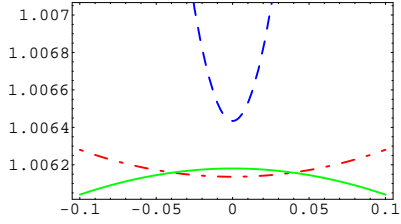}
\vspace*{-0.3cm}
\caption{\footnotesize {\boldmath$\Omega_{bc}/\Xi_{bc}:$} $b$ behaviour  of $r^{sd}_1(bc)$: dot-dashed line (red); $r^{sd}_2(bc)$: continuous line (green), and $r^{sd}_{12}(bc)$: dashed line (blue)  for $t_c=50$ GeV$^{2}$, $\tau=0.8$ GeV$^{-2}$, $m_c=1.26$ GeV and $m_b=4.22$ GeV.
}
\label{fig:lambda1}
\end{center}
\end{figure} 
\nin

   \vspace*{-0.5cm}
\begin{figure}[hbt]
\begin{center}
\includegraphics[width=5.2cm]{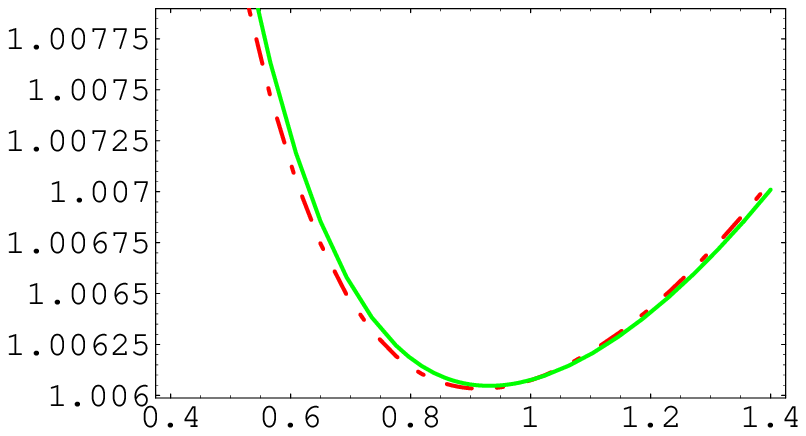}
\includegraphics[width=5.2cm]{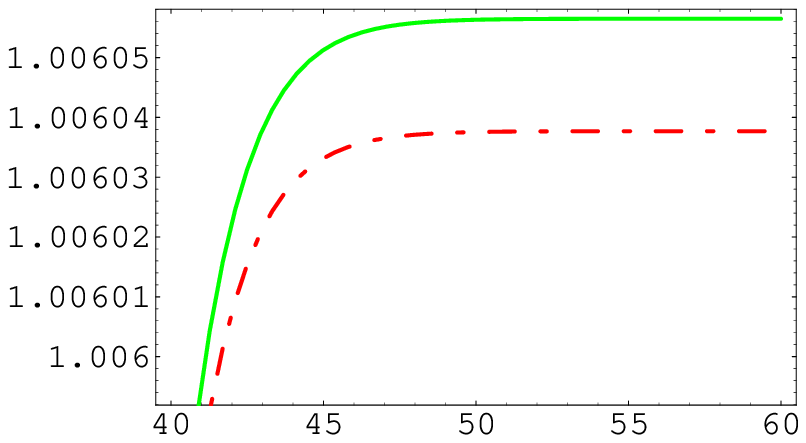}
\vspace*{-0.3cm}
\caption{\footnotesize {\boldmath$\Omega_{bc}/\Xi_{bc}$} a):  $\tau$-behaviour of $r^{sd}_1(bc)$: dot-dashed line (red) and $r^{sd}_2(bc)$: continuous line (green) for $k=-0.05$, $t_c=50$ GeV$^{2}$ and $m_c=1.26$ GeV. 
b):  $t_c$-behaviour of $r^{sd}_1(bc)$ and $r^{sd}_2(bc)$ for $\tau=0.9$ GeV$^{-2}$ and $k=-0.05$. }
\label{fig:lambda2}
\end{center}
\end{figure} 
\nin
The $\Xi(bc)$ and the $\Omega(bc)$ spin 1/2 baryons can be described
by the corresponding currents:
\bea
J_{\Xi_{bc}}&=&\epsilon_{\alpha\beta\lambda}\left[(c_\alpha^TC\gamma_5d_\beta)+k(c_\alpha^TCd_\beta)\gamma_5\right]
b_\lambda, \nnb\\
J_{\Omega_{bc}}&=&J_{\Lambda_{bc}}  ~~~(d\rar s)~,
\label{curl}
\eea
where $d,s$ are light quark fields, $c,b$ are heavy quark fields and $k$ is {\it a priori} an arbitrary mixing parameter. The expression of the corresponding two-point correlator has been obtained in the chiral limit $m_d=m_s=0$ by Refs. \cite{BAGAN,BC}. We have checked these expressions which we complete here by adding the $m_s$-corrections for the PT and quark condensate contributions. The expressions of these corrections are:
\bea
{\rm Im} F_1^{m_q}\vert _{pert}&=&-\frac{m_s m_c (1-k^2)}{128 \:\pi^3 \:t^2} \Bigg{[}
	  6 \:{\cal L}_1  \left[ (m_c^2 t^2 - 2 m_b^4 m_c^2) \right] \nnb\\
	&&+6 \:{\cal L}_2   m_c^2 t^2 
	- \lambda_{bc}  \Big{[} 2t^2 + \left( 5m_c^2 - 4m_b^2 \right)t \nnb\\
	&&- m_c^4
	+ 5m_b^2 m_c^2 + 2m_ b^4 \Big{]}
	\Bigg{]} \nnb\\	
{\rm Im} F_1^{m_s}\vert_ {\bar ss}&=&\frac{m_s \ss (1+k^2)}{32 \:\pi \: t^3}\Bigg{[}	
	 \lambda_{bc} [ t^2 + (m_b^2 + m_c^2)t \nnb\\
	 &&-2(m_b^2-m_c^2)^2 ]
	+ \frac{2}{\lambda_{bc}}  [ (m_b^4+m_c^4)t^2 \nnb\\ &&-2(m_b^2+m_c^2)(m_b^2-m_c^2)^2 t 
	+ (m_b^2 - m_c^2)^4 ] 		
	\Bigg{]},\nnb\\
\eea	

\bea
{\rm Im} F_2^{m_s} \vert_ {pert}&=&-\frac{3 m_s m_c m_b(1+k^2)}{64 \:\pi^3 t} \Big{[}
	  2 \:{\cal L}_1  [ (m_b^2+m_c^2) t - \nnb\\ 
	 && 2m_b^2 m_c^2 ]
	-2 \:{\cal L}_2   \left( m_b^2 - m_c^2 \right)t \nnb\\
	 &&-\lambda_{bc}  \left[ s + m_ b^2 + m_ c^2 \right]
	\Big{]}\nnb\\
{\rm Im F}_2^{m_s }\vert_{\bar ss}&=&\frac{m_s m_b \ss (1-k^2)}{16 \:\pi t^2} \Big{[} \lambda_{bc} \left( t-m_b^2+m_c^2 \right) + \nnb\\
&&\frac{1}{\lambda_{bc}} [ m_b^2 t^2
+ \left( m_c^4+m_b^2 m_c^2-2m_b^4 \right)t \nnb\\
&&+ \left( m_b^2 - m_c^2 \right)^3  ] \Big{]}~,
\eea
where:
\bea
&&v=\sqrt{1-{4m^2_bm_c^2\over (t-m_b^2-m_c^2)^2}},~~\lambda_{bc}^{1/2}=(t-m_b^2-m_c^2)v\nnb\\
&&{\cal L}_1={1\over 2}\log{1+v \over 1-v}\nnb\\
&&{\cal L}_2=\log{(m_b^2+m_c^2)t+(m_b^2-m_c^2)(\lambda_{bc}^{1/2}-m_b^2+m_c^2)\over 2m_bm_ct}.\nnb\\
\eea
Like in previous sections, we study the different ratios of moments in Figs. \ref{fig:lambda1} and  \ref{fig:lambda2}. As one can see in Fig.  \ref{fig:lambda1}a, $r^{sd}_1(bc)$ and $r^{sd}_2(bc)$ are quite stable in $k$ and present common solutions for :
\beq
k= \pm 0.05~,
\eeq
inside the range given in Eq. (\ref{eq:mixing2}), while $r^{sd}_{12}(bc)$ does not intersect with the other DRSR. The $\tau$ and $t_c$ behaviours given  in Fig. \ref{fig:lambda2}a,b are also very stable from which we deduce the DRSR:
 \beq
 r^{sd}(bc)\equiv {M_{\Omega_{bc}}\over M_{\Xi_{bc}}}=1.006(0.2)_{\bar ss}(1.4)_{m_s}(1)_{m_Q}~,
 \eeq
where the errors coming from other parameters are negligible. This implies:
 \beq
 M_{\Omega_{bc}}- M_{\Xi_{bc}}= 41(7) ~{\rm MeV}~,
\label{eq:lambdamass}
 \eeq
where we have used the QSSR central value  $M_{\Xi_{bc}}\simeq 6.86$ GeV in Eq. (\ref{eq:bc}). 
The size of the mass-splitting can be compared with the potential model prediction  about (70-89) MeV given in \cite{BC,BRAC}. 
 \vspace*{-0.5cm}
{\scriptsize
\begin{table}[hbt]
\setlength{\tabcolsep}{0.45pc}
 \caption{\scriptsize    QSSR predictions for the doubly heavy baryons mass ratios and splittings, which we compare with the Potential Model (PM) range of results in \cite{BC,BRAC}.  The PM prediction for the spin 3/2 is an average with the one for spin 1/2. The mass inputs are in GeV and the mass-splittings are in MeV.}
    {\footnotesize
\begin{tabular}{llll}
&\\
\hline
Mass ratios&Mass inputs&Mass plittings& PM   \\
\hline
${{\Xi^*_{cc}}/ {\Xi_{cc }}}= 1.0167(19)$&${\Xi_{cc }}=3.52$\cite{PDG}& ${{\Xi^*_{cc }}- {\Xi_{cc}}}=59(7)$&70-93\\
${{\Xi^*_{bb }}/ {\Xi_{bb}}}=1.0019(3)$&${\Xi_{bb}}=9.94$\cite{BAGAN}&${{\Xi^*_{bb }}- {\Xi_{bb}}}= 19(3)$&30-38\\
${{\Omega_{cc}}/ {\Xi_{cc}}}=1.0260(70)$&${\Xi_{cc }}=3.52$\cite{PDG}& ${\Omega_{cc}}- {\Xi_{cc}}= 92(24) $&90-102\\
${\Omega_{bb}}/ {\Xi_{bb}}=1.0049(13)$&${\Xi_{bb}}=9.94$\cite{BAGAN}&${\Omega_{bb}}- {\Xi_{bb}}= 49(13)$&60-73\\
${{\Omega^*_{cc}}/ {\Xi^*_{cc}}}=1.0260(75)$&${\Xi^*_{cc }}=3.58^{ ~*)}$&$ {\Omega^*_{cc}}- {\Xi^*_{cc}}= 94(27)$&91-100\\
${{\Omega^*_{bb}}/ {\Xi^*_{bb}}}=1.0050(15)$&${\Xi^*_{bb}}=9.96^{~ *)}$&${\Omega^*_{bb}}- {\Xi^*_{bb}}= 50(15)$&60-72\\
${{\Omega_{bc}}/ {\Xi_{bc}}}=1.0060(17)$&${\Xi_{bc}}=6.86$\cite{BC}&${\Omega_{bc}}- {\Xi_{bc}}= 41(7)$&70-89\\
\hline
\end{tabular}
\begin{quote}
\scriptsize$^{ *)}$ We have combined your results for the mass-splittings with the experimental value of 
$M_{\Xi_{cc}}$ and with the central value of $M_{\Xi_{bb}}$ in Eq. (\ref{eq:bagan}). 
\end{quote} 
}
\label{tab:summary}
\end{table}
}
\vspace*{-.5cm}
\nin
\section{Conclusions}
\nin
Our different results are summarized in Table \ref{tab:summary} and agree in most cases with the potential model predictions given in \cite{BC,RICHARD}:\\
-- The mass-splittings between the spin 3/2 and 1/2 baryons, derived in Eqs. (\ref{eq:chicc}) and (\ref{eq:chibb}) is essentially due to the radiative corrections in our approach and seems to behave like $1/M_Q$. \\
-- For the SU(3) mass-splittings, our results, derived in Eqs. (\ref{eq:omegac}) and (\ref{eq:omegab}) for the spin 1/2  and in Eqs. (\ref{eq:omega*c}) and (\ref{eq:omega*b}) for the spin 3/2,
indicate that the splittings due to the SU(3) breaking are almost independent on the spin of the heavy baryons but approximately behave like $1/M_Q$. These mass-behaviours can  be qualitatively understood from the QCD expressions of the corresponding correlators where the leading mass corrections behave like $m_s/m_Q$.  \\
-- Finally, we obtain, in Eq. (\ref{eq:lambdamass}), the SU(3) mass-splittings between the $\Omega(bcs)$ and $\Xi(bcd)$, which is about 1/2 of the potential model prediction. \\
 Our previous predictions can improved by including radiative corrections to the SU(3) breaking terms and can be tested, in the near future, at Tevatron and LHCb.
\section*{Acknowledgements}
\nin
We thank Marina Nielsen, Jean-Marc Richard and Valya Zakharov for some discussions. R.M.A acknowledges the LPTA-Montpellier for the hospitality where this work has been done. This work has 
been partly supported by the CNRS-IN2P3 within the Non-perturbative QCD program in Hadron Physics. 

\end{document}